\documentclass[iop]{emulateapj}

\usepackage{atbegshi}
\def\hackaltaffiltext#1#2{\AtBeginShipoutNext{\footnotetext[#1]{#2}\stepcounter{footnote}}}

\def\ltsima{$\; \buildrel < \over \sim \;$}
\def\simlt{\lower.5ex\hbox{\ltsima}}
\def\gtsima{$\; \buildrel > \over \sim \;$}
\def\simgt{\lower.5ex\hbox{\gtsima}}
\def\kms{{\rm\,km\,s^{-1}}}

\def\kpc{{\rm\,kpc}}

\def\msun{{\rm\,M_\odot}}
\def\lsun{{\rm\,L_\odot}}

\def\pc{{\rm\,pc}}

\def\deg{^\circ}

\def\s{\ifmmode \widetilde \else \~\fi}
\def\={\overline}

\def\spose#1{\hbox to 0pt{#1\hss}}

\def\lta{\mathrel{\spose{\lower 3pt\hbox{$\mathchar"218$}}
     \raise 2.0pt\hbox{$\mathchar"13C$}}}
\def\gta{\mathrel{\spose{\lower 3pt\hbox{$\mathchar"218$}}
     \raise 2.0pt\hbox{$\mathchar"13E$}}}
\def\Dt{\spose{\raise 1.5ex\hbox{\hskip3pt$\mathchar"201$}}}    
\def\dt{\spose{\raise 1.0ex\hbox{\hskip2pt$\mathchar"201$}}}    

\def\dotsfill{\leaders\hbox to 1em{\hss.\hss}\hfill}

\def\Gyr{{\rm\,Gyr}}
\def\FeH{{\rm[Fe/H]}}

\shorttitle{Hydra~II: a new Milky Way dwarf galaxy}
\shortauthors{N. F. Martin et al.}

\begin{document}


\title{Hydra II: a faint and compact Milky Way dwarf galaxy found in the Survey of the Magellanic Stellar History}


\author{Nicolas F. Martin$^{1,2}$, David L. Nidever$^3$, Gurtina Besla$^4$, Knut Olsen$^5$, Alistair R. Walker$^{6}$, A. Katherina Vivas$^{6}$, Robert A. Gruendl$^{7,8}$, Catherine C. Kaleida$^{5,6}$, Ricardo R. Mu\~noz$^{9,25}$, Robert D. Blum$^5$, Abhijit Saha$^5$, Blair C. Conn$^{10}$, Eric F. Bell$^3$, You-Hua Chu$^{11,8}$, Maria-Rosa L. Cioni$^{12,13,14}$, Thomas J. L. de Boer$^{15}$, Carme Gallart$^{16,17}$, Shoko Jin$^{18}$, Andrea Kunder$^{13}$, Steven R. Majewski$^{19}$, David Martinez-Delgado$^{20}$, Antonela Monachesi$^{21}$, Matteo Monelli$^{16,17}$, Lara Monteagudo$^{16,17}$, Noelia E. D. No\"el$^{22}$, Edward W. Olszewski$^4$, Guy S. Stringfellow$^{23}$, Roeland P. van der Marel$^{24}$, Dennis Zaritsky$^{4}$}

\email{nicolas.martin@astro.unistra.fr}

\altaffiltext{1}{Observatoire astronomique de Strasbourg, Universit\'e de Strasbourg, CNRS, UMR 7550, 11 rue de l'Universit\'e, F-67000 Strasbourg, France}
\altaffiltext{2}{Max-Planck-Institut f\"ur Astronomie, K\"onigstuhl 17, D-69117 Heidelberg, Germany}
\altaffiltext{3}{Department of Astronomy, University of Michigan, 1085 S. University Ave., Ann Arbor, MI 48109-1107, USA}
\altaffiltext{4}{Steward Observatory, University of Arizona, 933 North Cherry Avenue, Tucson AZ, 85721}
\altaffiltext{5}{National Optical Astronomy Observatory, 950 N Cherry Ave, Tucson, AZ 85719, USA} 
\altaffiltext{6}{Cerro Tololo Inter-American Observatory, National Optical Astronomy Observatory, Casilla 603, La Serena, Chile}
\altaffiltext{7}{National Center for Supercomputing Applications, 1205 West Clark St., Urbana, IL 61801, USA}
\altaffiltext{8}{Department of Astronomy, University of Illinois, 1002 West Green St., Urbana, IL 61801, USA}
\altaffiltext{9}{Departamento de Astronom\'ia, Universidad de Chile, Camino del Observatorio 1515, Las Condes, Santiago, Chile}
\altaffiltext{10}{Gemini Observatory, Recinto AURA, Colina El Pino s/n, La Serena, Chile.}
\altaffiltext{11}{Institute of Astronomy and Astrophysics, Academia Sinica, No.1, Sec. 4, Roosevelt Rd, Taipei 10617, Taiwan, R.O.C.}
\altaffiltext{12}{Universit\"{a}t Potsdam, Institut f\"{u}r Physik und Astronomie, Karl-Liebknecht-Str. 24/25, 14476 Potsdam, Germany}
\altaffiltext{13}{Leibniz-Institut f\"{u}r Astrophysics Potsdam (AIP), An der Sternwarte 16, 14482 Potsdam Germany}
\altaffiltext{14}{University of Hertfordshire, Physics Astronomy and Mathematics, Hatfield AL10 9AB, United Kingdom}
\altaffiltext{15}{Institute of Astronomy, University of Cambridge, Madingley Road, Cambridge CB3 0HA, UK}
\altaffiltext{16}{Instituto de Astrof\'{i}sica de Canarias, La Laguna, Tenerife, Spain}
\altaffiltext{17}{Departamento de Astrof\'{i}sica, Universidad de La Laguna, Tenerife, Spain}
\altaffiltext{18}{Kapteyn Astronomical Institute, University of Groningen, P.O. Box 800, 9700 AV Groningen, The Netherlands}
\altaffiltext{19}{Department of Astronomy, University of Virginia, Charlottesville, VA 22904, USA}
\altaffiltext{20}{Astronomisches Rechen-Institut, Zentrum f\"ur Astronomie der Universit\"at Heidelberg,  M{\"o}nchhofstr. 12-14, 69120 Heidelberg, Germany}
\hackaltaffiltext{21}{Max-Planck-Institut f\"ur Astrophysik, Karl-Schwarzschild-Str. 1, 85748 Garching, Germany}
\hackaltaffiltext{22}{Department of Physics, University of Surrey, Guildford, GU2 7XH, UK}
\hackaltaffiltext{23}{Center for Astrophysics and Space Astronomy, University of Colorado, 389 UCB, Boulder, CO, 80309-0389, USA}
\hackaltaffiltext{24}{Space Telescope Science Institute, 3700 San Martin Drive, Baltimore, MD 21218}
\hackaltaffiltext{25}{Visiting astronomer, Cerro Tololo Inter-American Observatory, National Optical Astronomy Observatory, which is operated by the Association of Universities for Research in Astronomy (AURA) under a cooperative agreement with the National Science Foundation.}

\begin{abstract}
We present the discovery of a new dwarf galaxy, Hydra~II, found serendipitously within the data from the ongoing Survey of the \textsc{Ma}gellanic Stellar History (SMASH) conducted with the Dark Energy Camera on the Blanco 4m Telescope. The new satellite is compact ($r_h = 68 \pm 11\pc$) and faint ($M_V = -4.8\pm0.3$), but well within the realm of dwarf galaxies. The stellar distribution of Hydra~ II in the color-magnitude diagram is well-described by a metal-poor ($\FeH=-2.2$) and old (13 \Gyr) isochrone and shows a distinct blue horizontal branch, some possible red clump stars, and faint stars that are suggestive of blue stragglers. At a heliocentric distance of $134\pm 10\kpc$, Hydra~II is located in a region of the Galactic halo that models have suggested may host material from the leading arm of the Magellanic Stream. A comparison with N-body simulations hints that the new dwarf galaxy could be or could have been a satellite of the Magellanic Clouds.
\end{abstract}


\section{Introduction}
Our view of the Milky Way satellite system has thoroughly changed over the last decade, thanks to systematic CCD surveys of large swathes of the sky. Explorations of the Sloan Digital Sky Survey (SDSS) first revealed that dwarf galaxies extend to a much fainter regime than previously thought \citep[e.g.,][]{willman05a} and more than doubled the number of known Milky Way dwarf galaxies from data mainly covering the North Galactic Cap \citep[e.g.,][]{belokurov07a}. Spectroscopic surveys \citep[e.g.,][]{munoz06b,martin07a,simon07} confirm the dwarf galaxy nature of these systems by demonstrating them to be kinematically hotter than implied solely by their baryonic content --- which can be as low as only a few hundred solar luminosities \citep{martin08b}.

More recently, a cohort of faint satellites, most of them likely dwarf galaxies, was found within the first year data from the Dark Energy Survey \citep{bechtol15,koposov15}. Although these data only cover 1,800~deg$^2$ of the southern Galactic cap to the north of the Magellanic Clouds, they revealed 9 new stellar systems. The majority of these new systems have distances that hint that they likely are, or were, associated with the Clouds.

Within the $\Lambda$CDM cosmology, the Milky Way is expected to grow hierarchically by accreting groups of smaller galaxies over time \citep[e.g.,][]{li07} and it is therefore not unexpected that the Large Magellanic Cloud (LMC) would be in the process of shedding its satellites in the Milky Way halo \citep{donghia08}. In fact, comparisons with simulations show that the majority of current Milky Way satellites, especially the fainter ones, could have been accreted as satellites of larger dwarf galaxies \citep{wetzel15}. These group accretions could then drive the planar distribution of Milky Way satellites \citep{lynden-bell76,metz07,pawlowski13a}, even though the flat spatial distribution of Milky Way satellites still proves challenging to reproduce in a $\Lambda$CDM universe (\citealt{pawlowski12}, but see also, e.g., \citealt{wang13}).

Here, we report the discovery of a new dwarf galaxy found in the data of the Survey of the \textsc{Ma}gellanic Stellar History (SMASH; PI D. Nidever). Located at $(\alpha,\delta) = (12^\mathrm{h}21^\mathrm{m}42.1^\mathrm{s}, -31\deg59'07'')$ in the Hydra constellation, we name it Hydra~II (HyaII) following the usual naming conventions\footnote{Hydra~I is a concentration of stars that has been proposed as the progenitor of the `East Banded Structure,' a stellar structure at the edge of the Milky Way disk \citep{grillmair11}.}.

Throughout this paper, the distance to the Milky Way center is assumed to be $8\kpc$.

\section{The SMASH data and discovery}
SMASH is a NOAO community survey using $\sim$40 nights with DECam on the CTIO Blanco 4m to perform deep imaging over $\sim$2,400 deg$^2$ of the southern sky (at 20\% filling factor) with the goal of studying the complex stellar structures of the Magellanic Clouds at all angular separations with unprecedented fidelity (D. Nidever et al. 2015, in preparation). This letter uses data taken during UT 2013 March 17--20 under partly photometric conditions. Observations were obtained for 23 fields spread across an area of $\sim$1,200 deg$^2$ in the region of the leading arm of the Magellanic Stream. The data used here consist of 1$\times$60s and 3$\times$267s in $gr$ and 3$\times$333s in $iz$ dithered subexposures. Observations of several standard star fields at various airmasses were also obtained every night to enable photometric calibration.

InstCal image data products (calibrated, single-frame images) are automatically produced by the DECam Community Pipeline (CP; \citealt{valdes14}) and provided by the NOAO Science Archive Server\footnote{\url{https://www.portal-nvo.noao.edu}}. The rest of the photometric reduction is performed with the PHOTRED\footnote{\url{https://github.com/dnidever/PHOTRED}} pipeline first described in \citet{nidever11}. PHOTRED is an automated PSF photometry pipeline based on the DAOPHOT suite of programs \citep{stetson87,stetson94} and performs WCS fitting, single-image PSF photometry (ALLSTAR), forced PSF photometry of multiple images with a master source list created from a deep stack of all exposures (ALLFRAME), aperture correction, photometric calibration, and dereddening. Finally, star-like detections are separated from extended sources by enforcing a cut on the Sextractor stellar probability index ($\textrm{prob}>0.8$).

HyaII was discovered through a search for compact overdensities in the 23 SMASH year 1 fields. As in \citet{koposov07}, the data are first filtered to isolate relatively blue stars ($-1.0<(g-r)_0<1.0$) compatible with a dwarf galaxy's red giant branch (RGB), horizontal branch (HB), old main-sequence turn-off (oMSTO), and main sequence before being binned in $0.5'$ pixels and convolved with a $1'$ Gaussian kernel. A visual inspection of the resulting maps reveals a few significant overdensities, most of which are spurious, triggered by known astronomical objects in the background. The only overdensity found without a counterpart in the NASA/IPAC Extragalactic Database or Simbad is HyaII.

\section{The Properties of Hydra~II}
\begin{figure*}
\begin{center}
\includegraphics[width=0.375\hsize,angle=270]{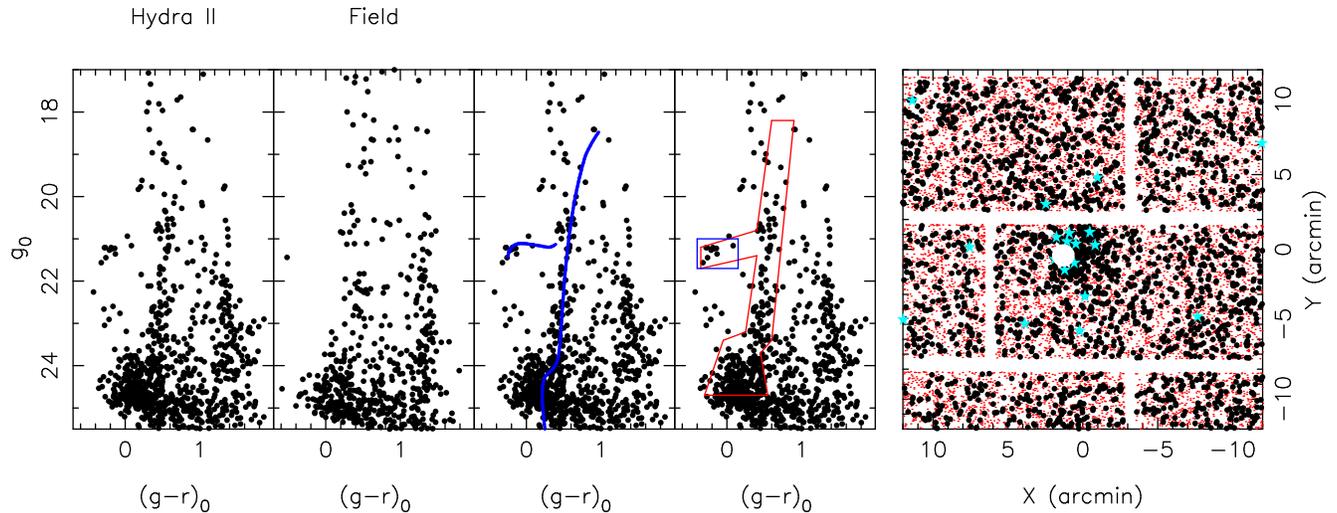}
\caption{\label{CMDs}\emph{Left:} CMD of stellar-like objects within 2 half-light radii of HyaII's centroid, that of a field comparison with the same area, the CMD of HyaII with an old and metal-poor isochrone overlaid, and the CMD of HyaII with our CMD selection boxes overlaid. The isochrone shown is taken from the \textsc{Parsec} library \citep{bressan12} and has an age of $13\Gyr$ and metallicity of $\FeH = -2.2$. \emph{Right:} Spatial distribution of stellar-like objects in and around HyaII. Small red dots represent all star-like objects in the SMASH catalogue, whereas black dots correspond to those selected along the CMD features of HyaII; cyan star symbols correspond to potential member HB stars (see text). The white stripes indicate the location of chip gaps and the white circle an exclusion region of unreliable photometry affected by a bright star.}
\end{center}
\end{figure*}

Figure~\ref{CMDs} presents the color-magnitude diagram (CMD) of HyaII for a region within 2 half-light radii ($3\farcm4$) of its centroid, as determined below. The features produced by HyaII become evident when compared to the field CMD shown in the same figure for a region of the same size and taken $18 '$ to the west. One can see the dwarf galaxy's almost vertical RGB between $[(g-r)_0,g_0] = [0.4,23.5]$ and $[(g-r)_0,g_0] = [0.7,20.0]$, where it fades into the foreground contamination, its very clear blue HB at $[(g-r)_0,g_0] =\sim [-0.2,21.2]$, its oMSTO and the beginning of the corresponding MS at $[(g-r)_0,g_0] \sim [0.3,24.0]$ and fainter. HyaII also hosts some stars that are consistent with being red clump stars at $[(g-r)_0,g_0] \sim [0.5,20.8]$. A handful of stars bluer and brighter than the oMSTO at $(g-r)_0\sim-0.2$ and $22.0<g_0<23.5$ are likely blue stragglers or intermediate age stars. The smaller relative number of stars in this group to those in the HyaII blue HB is consistent with ratios seen for blue straggler populations in faint dwarf galaxies \citep{deason15}.

The right-most panel of Figure~\ref{CMDs} shows the distribution of all stellar-like objects in this region of the SMASH survey as small red dots, whereas the big black dots correspond to stars specifically selected in the selection box shown in red in the right-most CMD panel of Figure~\ref{CMDs}. Despite the chip gaps and the presence of a bright star near the center of HyaII, the CMD features correspond to a clearly defined spatial overdensity that subtends a few arcminutes.

\begin{figure}
\begin{center}
\includegraphics[width=0.6\hsize,angle=270]{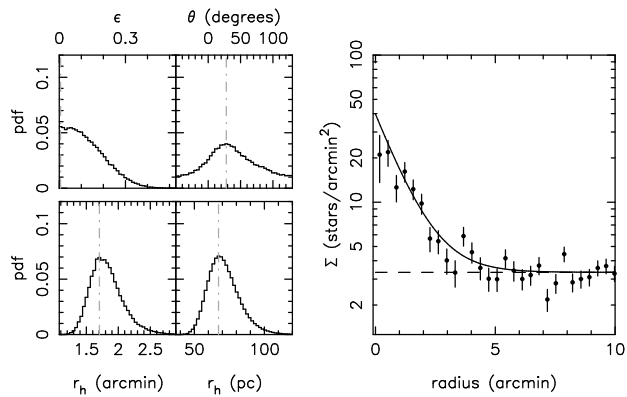}
\caption{\label{structure} \emph{Left:} One-dimensional, marginalized pdfs for the ellipticity ($\epsilon$), the position angle ($\theta$), and the angular and physical half-light radii ($r_h$) of HyaII. The gray dot-dashed lines indicate the modes of the distributions. \emph{Right:} Radial density profile of the favored exponential model (full line) compared to the data binned with the favored centroid, ellipticity, position angle, and number of stars within the chosen CMD selection box. The dashed line represents the inferred contamination level. The good match between the data and the model radial density profile illustrates the good quality of the structural parameter inference.}
\end{center}
\end{figure}

We apply the algorithm presented in an upcoming contribution (N. F. Martin et al., in preparation) to determine the structural parameters of HyaII. In summary, this algorithm builds on the work of \citet{martin08b} and uses the location of every star in the sample to produce Markov Chain Monte Carlo samplings of the posterior probability distribution functions (pdfs) for a family of models representing the spatial distribution of member stars as the sum of an exponential radial density profile and a constant level of contamination. The inference is performed in a 6-dimensional parameter space, with the parameters representing the centroid of the dwarf galaxy, its exponential half-light radius, $r_h$, its ellipticity, $\epsilon$, the position angle of its major axis East of North, $\theta$, and the number of stars in the galaxy for the chosen CMD selection box, $N^*$. This version of the algorithm further allows for the masking of regions without data, a particularly necessary step in the case of HyaII given the presence of the bright star near the dwarf galaxy's center. The resulting, marginalized pdfs are shown in Figure~\ref{structure} for the three most important structural parameters ($\epsilon$, $\theta$, and $r_h$) and summarized in Table~1. HyaII is a rather compact ($r_h=1\farcm7^{+0.3}_{-0.2}$) and round stellar system. However, the presence of the chip gaps and the necessity of masking out a region near the center of the dwarf galaxy render the structural parameter somewhat uncertain. Therefore, looking at the pdfs in Figure~\ref{structure}, it is still possible for HyaII to be elliptical ($\epsilon \simlt 0.4$) and larger than the favored model ($r_h\simlt 2\farcm5$).

The distinct HB of HyaII provides us with a good tool for inferring the distance to the dwarf galaxy. A small CMD selection box tailored around the HB (blue polygon overlaid on the right-most CMD of Figure~\ref{CMDs}) isolates stars shown in cyan in the right-most panel of Figure~\ref{CMDs}. Further requiring that they be within $4r_h$ limits the sample to 12 HB stars with high membership probability. The relation given by \citet[][their Equation~7]{deason11} between the absolute magnitude of blue HB stars and their colors allows us to calculate a distance modulus for each of these 12 stars, which, when combined, yield the average distance modulus for HyaII: $(m-M)_0 = 20.64\pm0.16$, where we assumed an uncertainty of 0.10 on the relation. This corresponds to a heliocentric distance of $134\pm10\kpc$, or a Galactocentric distance of $128\pm10\kpc$. This value further yields an estimate of the physical size of the system, $r_h = 68\pm11\pc$ (folding in the uncertainties on the distance), with the corresponding pdf presented in the bottom-right pdf panel of Figure~\ref{structure}.

On the third CMD of Figure~\ref{CMDs}, we have plotted an old ($13\Gyr$) and metal-poor ($\FeH=-2.2$) \textsc{Parsec} isochrone \citep{bressan12}. The isochrone has been shifted to the distance modulus of 20.64 determined above and nicely tracks all the CMD features of the dwarf galaxy (HB, RGB, MSTO).

Following Martin et al. (in preparation) and \citet{martin08b}, we use the inference of $N^*$, the number of HyaII stars within a chosen CMD selection box, corrected for holes in the data, to determine the absolute magnitude of HyaII. Although this means that we do not directly calculate the luminosity contribution of the observed stars, the resulting quantity is more representative of the underlying properties of dwarf galaxies, folding in `CMD shot noise' \citep{martin08b}. We first rerun our structural parameter code for a shallower CMD selection box ($g_0<23.7$), for which the data are complete. We then generate an artificial CMD from the isochrone shown above and its associated luminosity function (assuming a \citealt{kroupa01} IMF), folding in the photometric uncertainties determined from the data of the SMASH field. From the shallow structural parameter chain, we randomly draw a value $N^*_i$ of the number of stars we are aiming for and then populate the artificial CMD until $N^*_i$ stars have been drawn in the shallow CMD box. Summing up the flux of all stars of the artificial CMD, whether in the CMD box or not, yields the total luminosity of the dwarf galaxy. In addition, reiterating this procedure for different values of $N^*_i$ drawn from its pdf allows us to track the influence of `CMD shot noise' on our uncertainties. We infer a total luminosity of $L_V=10^{3.9\pm0.1}\lsun$, i.e., $M_V = -4.8\pm 0.3$.

\begin{table}
\caption{\label{properties}Properties of Hydra~II}
\begin{tabular}{c|c}
\hline
$\alpha$ (J2000) & $12^\mathrm{h}21^\mathrm{m}42.1^\mathrm{s}$\\
$\delta$ (J2000) & $-31\deg59'07''$\\
$\ell$ & $295.61\deg$\\
$b$ & $+30.46\deg$\\
$(m-M)_0$ & $20.64\pm0.16$\\
Heliocentric Distance & $134\pm10\kpc$\\
Galactocentric Distance & $128\pm10\kpc$\\
$M_{V}$ & $-4.8\pm 0.3$\\
$L_{V}$ & $10^{3.9\pm0.1}\lsun$\\
$E(B-V)^{a}$ & 0.053\\
Ellipticity & $0.01^{+0.19}_{-0.01}$\\
Position angle (E of N) & $28^{+40}_{-35}$$\deg$\\
$r_{h}$ & $1\farcm7^{+0.3}_{-0.2}$\\
& $68\pm11\pc$\\
$^a$ from \citet{schlafly11} &\\
\end{tabular}
\end{table}

\section{Discussion}

\begin{figure*}
\begin{center}
\includegraphics[width=0.5\hsize,angle=270]{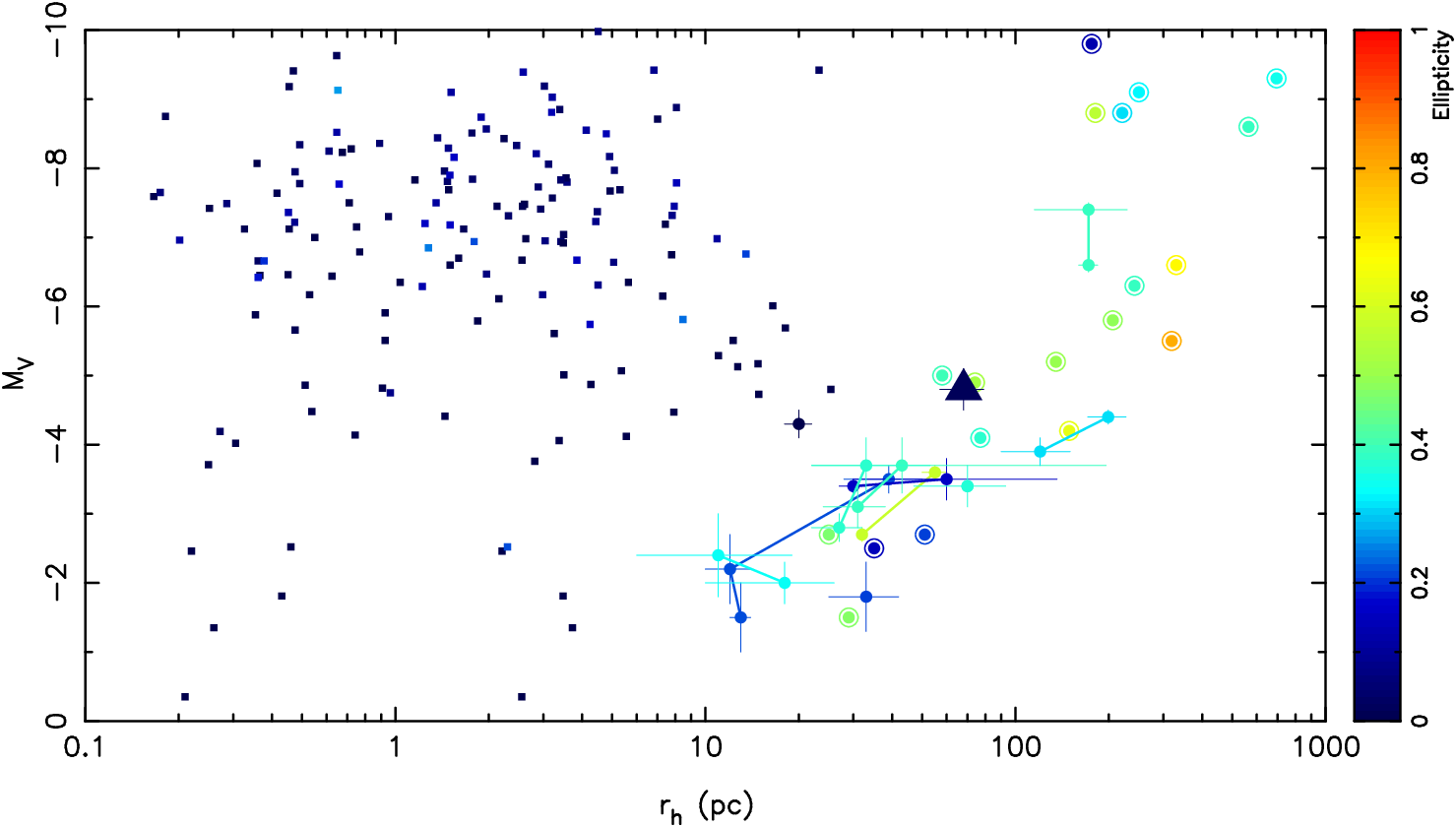}
\caption{\label{rh_Mv} Distribution of Milky Way satellites in the $r_h$--$M_V$ plane, color-coded by ellipticity. Squares represent globular clusters \citep{harris96,harris10}, circled dots are confirmed dwarf galaxies \citep{mcconnachie12}, while dots are recently discovered and ambiguous systems for which there are no spectroscopic confirmations as to their nature. These include the DES discoveries \citep{bechtol15,kim15,koposov15}, PSO~J174.0675-10.8774/Crater \citep{belokurov14,laevens14}, and Laevens~2/Triangulum~II \citep{laevens15a}. Symbols connected by a full line correspond to different measurements of the size and magnitude of the same object. HyaII is represented by the large triangle and falls within the realm of known, faint dwarf galaxies.}
\end{center}
\end{figure*}

The properties of the new dwarf galaxy, Hydra~II, are typical of other faint dwarf galaxies found around the Milky Way, as can be seen in Figure~\ref{rh_Mv}. HyaII is compact ($r_h=68\pm11\pc$) and faint ($M_V = -4.8\pm0.3$) but shares these properties with Coma Berenices \citep{munoz10} and Canes Venatici~II \citep{sand12}. HyaII is surprisingly round for a dwarf galaxy but it should be noted that this parameter is not very well constrained given the presence of a bright star near the center of the dwarf galaxy. Finally, the oldest and most metal-poor stellar population present in the \textsc{Parsec} library provides a good fit to the CMD features of HyaII, as is often the case for such faint dwarf galaxies \citep[e.g.,][]{brown14}.

\begin{figure}
\begin{center}
\includegraphics[width=\hsize]{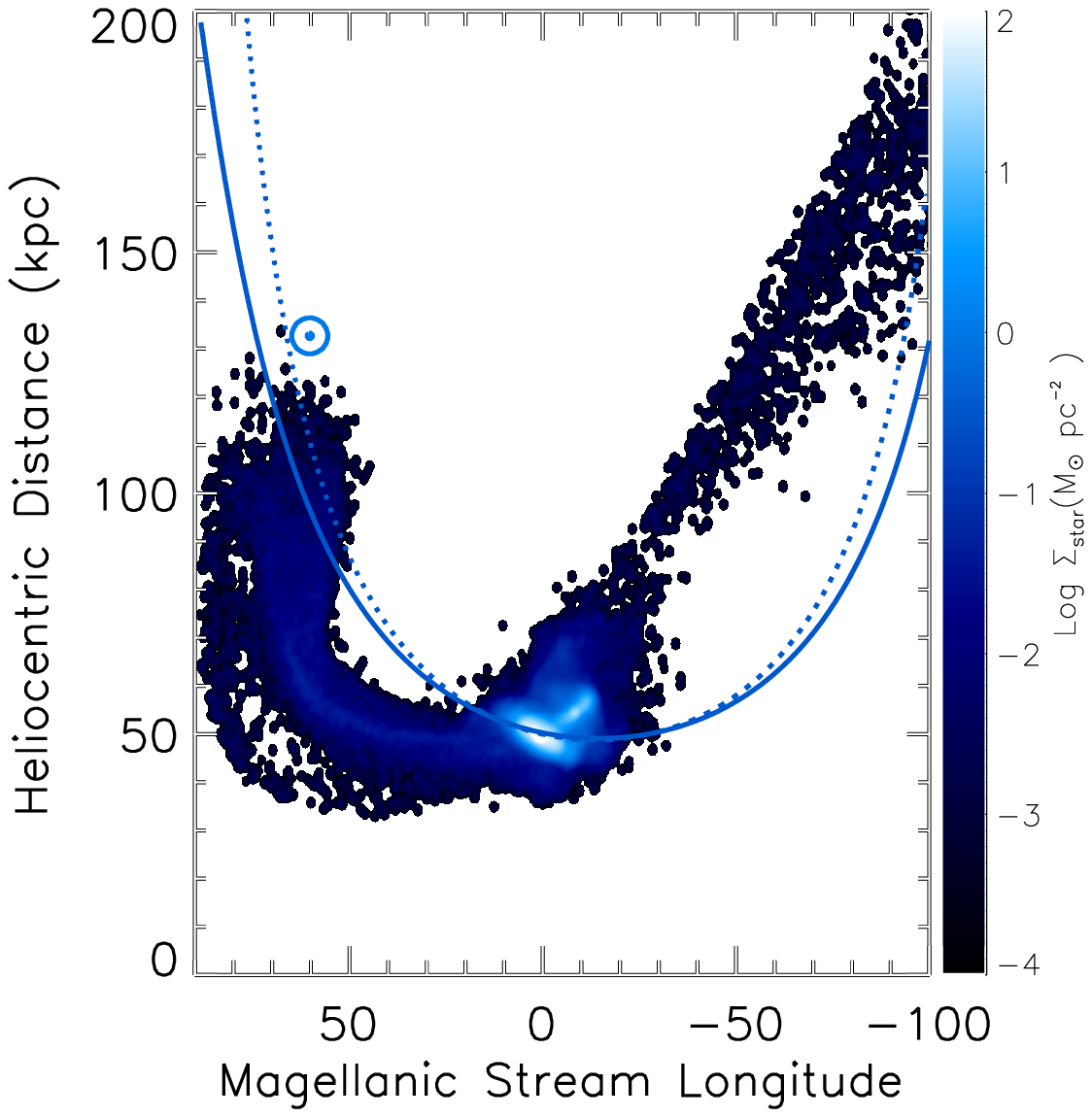}
\caption{\label{MC_Comparison}Heliocentric distances to simulated stellar debris stripped from the SMC by LMC tides as a function of Magellanic Stream Longitude (MSL), as defined in \citet{nidever08}. Simulation data are taken from Model~2 of \citet{besla12}. Color coding indicates stellar surface density. The LMC and SMC are located near $\textrm{MSL}=0\deg$ and are currently moving towards positive MSL. The past and future orbital trajectories of the Clouds in a static $10^{12}\msun$ Milky Way model are determined using two different velocities for the LMC: the dashed blue line corresponds to an LMC velocity of 380 km/s \citep{kallivayalil06}, which was also used to create this simulation, while the solid blue line denotes the orbit calculated with the lower velocity measured from three epochs of HST data \cite[$320\kms$;][]{kallivayalil13}. The revised speeds do not markedly change the future orbital path of the Clouds. The location of HyaII is marked by the $\odot$ symbol and is located at a distance consistent with debris removed from the SMC. It should nevertheless be noted that HyaII resides at a Magellanic Stream Latitude of $-18\deg$ and is not exactly on the Magellanic Stream plane.}
\end{center}
\end{figure}

HyaII is located close to the Vast Polar Structure of Milky Way satellites proposed by \citet{pawlowski12b} but this is hardly surprising given that the SMASH survey mainly targets fields in this proposed structure. HyaII is also located in the same region of the sky as the leading arm of the gaseous Magellanic Stream \citep{nidever10}, suggesting that HyaII is, or was until recently, part of the CloudsÕ satellite system. The orbital properties of possible satellites of the LMC are presently unconstrained, except for one, the Small Magellanic Cloud (SMC; \citealt{kallivayalil13}), believed to have been accreted by the Milky Way as part of the LMC group, which explains the SMCÕs proximity to the LMC, the gaseous bridge connecting them, and similarities in their star formation histories. It is thus reasonable to examine the orbital properties of this one confirmed companion to see if it can shed light on a plausible association of HyaII with the LMC. The orbit of the SMC is expected to be highly eccentric, explaining why the two Clouds have not yet merged. The SMC has likely completed 2--3 orbits about the LMC \citep{besla10}. At each pericentric approach between the two galaxies, LMC tides strip stars and gas from the SMC. Indeed, \citet{olsen11} and \citet{noel13} report a population of SMC stars stripped and captured by the LMC, and \citet{nidever13} found a stellar component of the gaseous Magellanic Bridge tidally stripped from the SMC. In Figure~\ref{MC_Comparison}, we examine the heliocentric distances to stellar debris stripped from the SMC as predicted by simulations that reproduce the global properties of the Magellanic System in a first infall scenario \citep[Model~2]{besla12}.

HyaII is located near the Leading Arm, a collection of high-velocity neutral hydrogen gas that leads the Magellanic Clouds. This structure is one of the main lines of evidence that the extended gas distribution surrounding the Clouds resulted from tidal forces rather than hydrodynamic processes, such as ram-pressure stripping. The latter process cannot pull material ahead of the stripped body, while the former process can naturally throw stripped material ahead of the Clouds. The Leading Arm is expected to have formed as a tidal tail over the course of the most recently completed orbit of the SMC about the LMC, whereas the Stream formed as a tidal tail at an earlier encounter. Predicted stellar debris in the Leading Arm should roughly trace the most recent orbit of the SMC about the LMC \citep{besla12}. Once removed, stripped SMC stars gain energy and migrate to larger radii than the original orbital path of the SMC about the LMC. In addition, Milky Way tides move the debris to even larger distances.  As a result, the distribution of these debris can reach distances as large as $130\kpc$ ($80\kpc$ from the LMC itself).

The distance to HyaII is consistent with the debris trail and trajectory of the LMC. The new dwarf galaxy is also within the virial radius of the LMC\footnote{It is, however, outside the LMC's tidal radius of $22 \pm 5$ kpc \citep{vandermarel14}.}, assuming a halo mass of $\sim10^{11}\msun$ as predicted by abundance-matching relations \citep{moster13}. \citet{sales11} demonstrated that most subhalos of a model-analogous LMC group should remain within the dark-matter halo of the LMC in a first infall scenario.  While not a definite proof, our analysis does suggest that the distance of HyaII is consistent with an LMC association. This study further suggests two clear tests for establishing the association of HyaII with the LMC: 
 \begin{itemize}
 \item The detection of stars in the Leading Arm that have metallicities consistent with a stripped SMC population can inform us of the extreme distances that material in orbit about the LMC can reach.  The identification of such stars is a major goal of our SMASH program.
 \item With radial velocities and proper motion measurements of HyaII's stars, the dwarf galaxy's 3D velocity can be compared against that of the Clouds to test whether the dwarf galaxy is moving in a similar orbital plane.
 \end{itemize}

\acknowledgments

DLN was supported by a McLaughlin Fellowship at the University of Michigan. EFB. acknowledges support from NSF grant AST 1008342. M-RC acknowledges support by the German Academic Exchange Service (DAAD). TdB acknowledges financial support from the ERC under Grant Agreement n. 308024. SJ is supported by the Netherlands Organization for Scientific Research (NWO) Veni grant 639.041.131. DM-D acknowledge support by Sonderforschungsbereich (SFB) 881 ``The Milky Way System'' of the German Research Foundation (DFB). RRM acknowledges partial support from CONICYT Anillo project ACT-1122 and project BASAL PFB-$06$. GSS is supported by grants from NASA.

Based on observations at Cerro Tololo Inter-American Observatory, National Optical Astronomy Observatory (NOAO Prop. ID: 2013B-0440; PI: Nidever), which is operated by the Association of Universities for Research in Astronomy (AURA) under a cooperative agreement with the National Science Foundation. This project used data obtained with the Dark Energy Camera (DECam), which was constructed by the Dark Energy Survey (DES) collaborating institutions: Argonne National Lab, University of California Santa Cruz, University of Cambridge, Centro de Investigaciones Energeticas, Medioambientales y Tecnologicas-Madrid, University of Chicago, University College London, DES-Brazil consortium, University of Edinburgh, ETH-Zurich, University of Illinois at Urbana-Champaign, Institut de Ciencies de l'Espai, Institut de Fisica d'Altes Energies, Lawrence Berkeley National Lab, Ludwig-Maximilians Universit\"at, University of Michigan, National Optical Astronomy Observatory, University of Nottingham, Ohio State University, University of Pennsylvania, University of Portsmouth, SLAC National Lab, Stanford University, University of Sussex, and Texas A\&M University. Funding for DES, including DECam, has been provided by the U.S. Department of Energy, National Science Foundation, Ministry of Education and Science (Spain), Science and Technology Facilities Council (UK), Higher Education Funding Council (England), National Center for Supercomputing Applications, Kavli Institute for Cosmological Physics, Financiadora de Estudos e Projetos, Funda\c{c}\~ao Carlos Chagas Filho de Amparo a Pesquisa, Conselho Nacional de Desenvolvimento Cient'fico e Tecnol\'ogico and the Minist\'erio da Ci\^encia e Tecnologia (Brazil), the German Research Foundation-sponsored cluster of excellence "Origin and Structure of the Universe" and the DES collaborating institutions.



\end{document}